# SEAL: Spatio-Textual Similarity Search

Ju Fan    Guoliang Li    Lizhu Zhou    Shanshan Chen    Jun Hu
Department of Computer Science and Technology, Tsinghua National Laboratory for Information Science and Technology (TNList), Tsinghua University, Beijing 100084, China.
fan-j07@mails.tsinghua.edu.cn, liguoliang@tsinghua.edu.cn,
dcszlz@tsinghua.edu.cn, aarifc33@gmail.com, j-hu08@mails.tsinghua.edu.cn

## ABSTRACT

Location-based services (LBS) have become more and more ubiquitous recently. Existing methods focus on finding relevant points-of-interest (POIs) based on users' locations and query keywords. Nowadays, modern LBS applications generate a new kind of spatio-textual data, *regions-of-interest* (ROIs), containing region-based spatial information and textual description, e.g., mobile user profiles with active regions and interest tags. To satisfy search requirements on ROIs, we study a new research problem, called spatio-textual similarity search: Given a set of ROIs and a query ROI, we find the similar ROIs by considering spatial overlap and textual similarity. Spatio-textual similarity search has many important applications, e.g., social marketing in location-aware social networks. It calls for an efficient search method to support large scales of spatio-textual data in LBS systems. To this end, we introduce a filter-and-verification framework to compute the answers. In the filter step, we generate signatures for the ROIs and the query, and utilize the signatures to generate candidates whose signatures are similar to that of the query. In the verification step, we verify the candidates and identify the final answers. To achieve high performance, we generate effective high-quality signatures, and devise efficient filtering algorithms as well as pruning techniques. Experimental results on real and synthetic datasets show that our method achieves high performance.

## 1. INTRODUCTION

Nowadays, as mobile devices (e.g., smartphones) with built-in global position systems (GPS) become more and more popular, location-based services (LBS) have been widely accepted by mobile users and attracted significant attention from both the academic and industrial community. Many location-based services, such as Foursquare[1] and Facebook Places[2], bring unique location-aware experiences to users.

[1] http://foursquare.com
[2] http://www.facebook.com/placesband

Existing LBS systems employ a spatial keyword search approach to provide LBS services [9, 7], which, given a set of points of interest (POIs), and a user query with location and keywords, finds all relevant POIs. For example, if a user wants to find the coffee shops nearby, she can issue a keyword query "`coffee shop`" to an LBS system, which returns the relevant coffee shops by considering the user's location and query keywords.

Recently, many modern LBS applications generate a new kind of spatio-textual data, *regions-of-interest* (ROIs), containing region-based spatial information and textual description. For example, in Facebook Places, mobile users have profiles consisting of active regions and interest tags. In wildlife monitoring, wild species with their habitats and descriptive features can be modeled by ROIs.

To satisfy search requirements on ROIs, we introduce a new research problem, called spatio-textual similarity search in this paper: Given a set of ROIs and a query ROI, we aim to find the ROIs which are similar to the query by considering spatial overlap and textual similarity.

Spatio-textual similarity search can satisfy users' information needs in various real applications. The first one is location-based social marketing using Facebook Places. As mentioned above, in Facebook Places, mobile users have profiles that can be modeled by ROIs. For example, consider some users in `Manhattan` who are interested in `tea` and `coffee`. A coffee shop (e.g., `starbucks`) can utilize user profiles in Facebook to provide location-specific advertisements to the potential customers who not only are interested in its products (e.g., {`starbucks`, `mocha`, `coffee`}) but also have region-based spatial overlap with its service area. Another example is friend recommendation in location-aware social networks, e.g., Facebook, Foursquare and Twitter[3]. Spatio-textual similarity search helps mobile users find potential friends with common interests (e.g., `playing basketball`) and overlap regions (e.g., `Brooklyn`), and thus facilitates users to form various kinds of *circles* with the same interests, such as sport games, shopping, and fans' activities. Spatio-textual similarity search can also support other applications, e.g., wildlife protection. Wild species have their habitats (e.g., `Yellowstone National Park` for grizzly bears) and features (e.g., `mammal`, `omnivore`, etc.). A zoologist can issue a query to find all wild species having certain features (e.g., `mammal`) and inhabiting in a specific region (e.g., `Idaho`).

In this paper, we formalize the problem of spatio-textual similarity search, and study the research challenges that naturally arise in this problem. A challenge is how to evaluate the similarity between two ROIs. Another challenge is how to achieve high search efficiency as LBS systems are required to support millions of users and respond to queries in milliseconds. Given a query ROI, there may be a huge amount of ROIs having significant overlaps with the query, thus it is rather expensive to find similar answers. Take the real dataset Twitter in our experiments (See Section 6) as an example. We used a set of query regions in an average area of 0.4 square kilometers. Even for one of the small query regions, there were, on average, 8000 ROIs overlapping with it. Moreover, similarity search needs to consider both spatial and textual similarities. To address these challenges, we propose an efficient Spatio-tExtuAl simiLarity search method, called SEAL-Search. We combine spatial similarity functions and textual similarity functions to quantify the similarity between two ROIs. To provide high performance, we introduce a filter-and-verification framework to compute the answers. In the filter step, our method generates *signatures* for spatio-textual objects and queries, and utilizes the signatures to generate candidates whose signatures are similar to those of the queries. In the verification step, it verifies the candidates and identifies the final answers. We develop effective techniques to generate signatures and devise efficient filtering algorithms to prune dissimilar objects.

[3] http://www.twitter.com

To summarize, we make the following contributions.

- To the best of our knowledge, we are the first to study spatio-textual similarity search on ROIs. We propose a filter-and-verification framework and signature-based filtering algorithms to address this problem.
- For effective spatial pruning, we devise grid-based signatures and develop threshold-aware pruning techniques.
- To utilize spatial and textual pruning simultaneously, we judiciously select high-quality signatures and devise efficient hybrid filtering algorithms.
- We have conducted extensive experiments on real and synthetic datasets. Experimental results show that our algorithms achieve high performance.

The paper is organized as follows. The problem formulation and related works are presented in Section 2. We introduce a signature-based method in Section 3. We develop grid-based filtering algorithms in Section 4 and hybrid filtering algorithms in Section 5. Experimental results are provided in Section 6. Finally, we conclude the paper and discuss the future work in Section 7.

## 2. PRELIMINARIES

### 2.1 Problem Formulation

**Data Model.** Our work focuses on supporting similarity search for a set of *spatio-textual ROI objects* (or objects for simplicity), $\mathcal{O} = \{o_1, o_2, \ldots o_{|\mathcal{O}|}\}$. Each object $o \in \mathcal{O}$ consists of spatial information $o.R$ and textual information $o.T$, denoted by $o = (R, T)$. The spatial information $o.R$ is a region. We use the well-known minimum bounding rectangle (MBR) to represent region $o.R$ through the bottom-left point and top-right point of the MBR. The textual description $o.T$ is a set of tokens, i.e., $\{t_1, t_2, \ldots, t_{|o.T|}\}$, where each token $t \in o.T$ is associated with a weight $w(t)$ to capture its importance. Figure 1 illustrates an example of seven objects, each of which has several tokens and a region.

**Query Model.** Our paper considers a spatio-textual similarity search query $q$ that also consists of a region $q.R$ and a

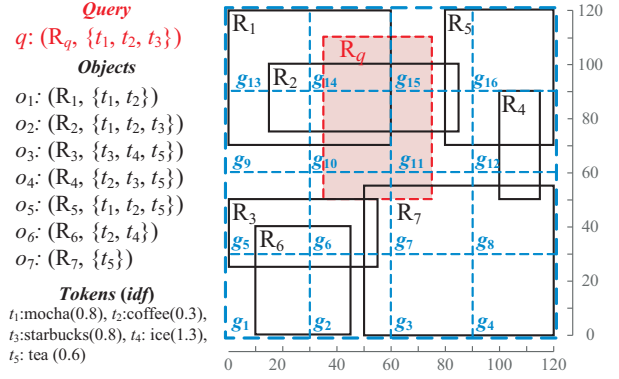

**Figure 1: An example of spatio-textual similarity search with objects $\{o_1, o_2, \ldots, o_7\}$ and query $q$.**

set of tokens $q.T = \{t_1, t_2, \ldots, t_{|q.T|}\}$. Given a set of objects $\mathcal{O} = \{o_1, o_2, \ldots, o_{|\mathcal{O}|}\}$, the answers of query $q$ are a set of objects $\mathcal{A} \subseteq \mathcal{O}$ *similar* to $q$, i.e., for $o \in \mathcal{A}$,

(1) The *Spatial Similarity* $\text{sim}_\text{R}(q, o) \geq \tau_\text{R}$, and
(2) The *Textual Similarity* $\text{sim}_\text{T}(q, o) \geq \tau_\text{T}$,

where $\tau_\text{R}$ and $\tau_\text{T}$ are respectively spatial and textual similarity thresholds satisfying $0 \leq \tau_\text{R}, \tau_\text{T} \leq 1$. Notice that we use the two thresholds to allow users to determine the spatial relevance and textual relevance in a more flexible way.

We quantify the spatial similarity based on the *overlap* of regions. Given two regions $q.R$ and $o.R$, their *overlap* is formally defined as the area of the intersecting region of $q.R$ and $o.R$, denoted by $|q.R \cap o.R|$. Note that we use operator $|\cdot|$ to represent both the cardinality of a set and the area of a region, if there is no ambiguity. Based on the overlap, we consider the Jaccard similarity in this paper.

DEFINITION 1 (SPATIAL SIMILARITY). *The spatial Jaccard similarity for two regions $q.R$ and $o.R$ is defined as*

$$\text{sim}_\text{R}(q, o) = \frac{|q.R \cap o.R|}{|q.R \cup o.R|},$$

*where $|q.R \cup o.R| = |q.R| + |o.R| - |q.R \cap o.R|$.*

For example, the overlap of regions $o_1.R$ and $q.R$ in Figure 1 is $|q.R \cap o_1.R| = 1000$ and $|q.R \cup o_1.R| = 4400$. Thus the spatial similarity is $\text{sim}_\text{R}(q, o_1) = 1000/4400 = 0.23$. Note that our method can be easily extended to other overlap-based functions, such as Dice Similarity.

On the other hand, textual similarity measures the similarity between two token sets $q.T$ and $o.T$. Many token-set based similarity functions have been studied in string similarity join/search [5, 18, 21], such as Jaccard similarity, Dice similarity, Cosine similarity, etc. In this paper, we take the weighted Jaccard similarity as an example.

DEFINITION 2 (TEXTUAL SIMILARITY). *The textual similarity $\text{sim}_\text{T}(q, o)$ between $q.T$ and $o.T$ is defined as the weighted Jaccard coefficient of the two token sets, i.e.,*

$$\text{sim}_\text{T}(q, o) = \frac{\sum_{t \in q.T \cap o.T} w(t)}{\sum_{t \in q.T \cup o.T} w(t)},$$

*where $w(t)$ is the weight of token $t$.*

In this paper we use the inverted document frequency (denoted by $\text{idf}$) as token weight, i.e., $w(t) = \ln \frac{|\mathcal{O}|}{\text{count}(t, \mathcal{O})}$, where $\text{count}(t,\mathcal{O})$ is the number of objects containing token $t$. Figure 1 shows five distinct tokens (i.e., $t_1 \sim t_5$) with their weights. Using the weights, we compute textual similarity $\text{sim}_\text{T}(q,o_1) = \big(w(t_1)+w(t_2)\big)/\big(w(t_1)+w(t_2)+w(t_3)\big) = 0.58$.

Based on the above-mentioned notations, we formalize the spatio-textual similarity search problem.

DEFINITION 3 (SPATIO-TEXTUAL SIMILARITY SEARCH). *Consider a set of spatio-textual objects $\mathcal{O} = \{o_1, o_2, \ldots, o_{|\mathcal{O}|}\}$ and a spatio-textual similarity search query $q = (R, T, \tau_\text{R}, \tau_\text{T})$. It returns the objects $\mathcal{A} \subseteq \mathcal{O}$ such that spatial similarity $\text{sim}_\text{R}(q,o) \geq \tau_\text{R}$ and textual similarity $\text{sim}_\text{T}(q,o) \geq \tau_\text{T}$, i.e., $\mathcal{A} = \{o \mid o \in \mathcal{O}, \text{sim}_\text{R}(q,o) \geq \tau_\text{R}, \text{sim}_\text{T}(q,o) \geq \tau_\text{T}\}$.*

EXAMPLE 1. *Consider the objects $\{o_1, o_2, \ldots, o_7\}$ and a query $q = (R_q, \{t_1, t_2, t_3\}, 0.25, 0.3)$ in Figure 1. Object $o_2 = (R_2, \{t_1, t_2, t_3\})$ is an answer since it satisfies $\text{sim}_\text{R} = 0.32 \geq \tau_\text{R}(0.25)$ and $\text{sim}_\text{T} = 1 \geq \tau_\text{T}(0.3)$. In contrast, object $o_1 = (R_1, \{t_1, t_2\},)$ is not an answer due to $\text{sim}_\text{R} = 0.23 < \tau_\text{R}$, although satisfying $\text{sim}_\text{T} = 0.58 \geq \tau_\text{T}$. Considering all the objects in $\mathcal{O}$, we obtain the answer of $q$, $\mathcal{A} = \{o_2\}$.*

## 2.2 Related Work

**Spatial Keyword Search:** There are many studies on spatial keyword search [25, 6, 11, 9, 23, 7, 24, 22, 3, 20, 4, 16, 14, 13]. One problem is knn based keyword search, which, given a query consisting of a location and a set of keywords, finds top-$k$ relevant POIs by considering distance and textual relevance. Felipe et al. [9] integrated signature files into R-tree, and Cong et al. [7] combined inverted files and R-tree. Another problem is region-based keyword search, which, given a query consisting of a region and a set of keywords, finds the relevant POIs relevant to the keywords in the region. The methods addressing the problem also employed the R-tree index, and integrated inverted lists of keywords into R-tree nodes [25, 11, 6].

Our spatio-textual similarity search problem is substantially different from the above-mentioned problems. The underlying data is a set of spatio-textual objects consisting of regions and tokens (i.e., ROIs), rather than POIs. Moreover, the query model is different, and we focus on spatio-textual similarity between objects and queries and devise efficient filtering algorithms for similarity search.

**String Similarity Search/Join:** The problem of string similarity search/join has been extensively studied [10, 1, 2, 5, 18, 19, 12]. Given a set of strings and a query string, string similarity search finds the strings whose similarities to the query are not smaller than a threshold $\tau$. Existing studies employed various functions, e.g., edit distance and Jaccard similarity, to quantify the similarity. To improve the performance, Chaudhuri et. al [5] proposed a prefix filtering framework, and Bayardo et al. [2] employed this framework to support Jaccard or Cosine similarity functions.

The basic idea of prefix filtering is to estimate a similarity upper bound of two sets using their subsets. Consider a string object $o$ and a query string $q$. The prefix filtering framework first maps both strings to sets, denoted by $S(o)$ and $S(q)$, and transforms various similarity functions to the overlap similarity on sets. More formally, if $\text{sim}(q,o) \geq \tau$, then the sets satisfy $|S(q) \cap S(o)| \geq c$, where $c$ is a threshold deduced from $\tau$. Then, the framework fixes a global order on the elements of all sets, and sorts the elements in each set based on the global order. Let $S^p(o)$ denote the prefix of set $S(o)$ consisting of the first $|S(o)| - c + 1$ elements. It is easy to prove that if $|S(q) \cap S(o)| \geq c$, their prefix sets must have overlap, i.e., $S^p(q) \cap S^p(o) \neq \emptyset$. Therefore, we can prune the dissimilar objects satisfying $S^p(q) \cap S^p(o) = \emptyset$.

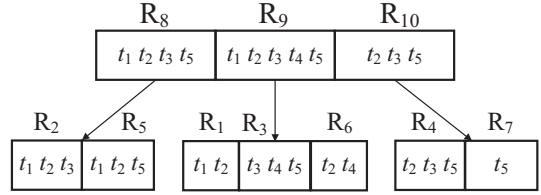

**Figure 2: The IR-tree index of objects in Figure 1.**

We utilize the prefix filtering framework to prune dissimilar objects. Compared with existing methods, our method focuses on devising effective threshold-aware pruning based on judiciously selected spatio-textual signatures.

**Grid-Based Spatial Index Structures:** The grid-based spatial indexes, such as Grid File and EXCELL, have been studied in spatial databases [17]. These methods decomposed the underlying space into a set of grids, and stored POIs into grids for fast access. We utilize grids in a different way: We employ grids as signatures of spatio-textual objects, and develop efficient spatial pruning techniques.

## 2.3 Baseline Methods

We introduce several straightforward methods to address the problem defined above, and will show their poor performance using experimental results (see Section 6).

**Keyword-first method.** The method constructs inverted indexes by mapping tokens to objects containing the tokens. Given a query, it first finds the objects with $\text{sim}_\text{T} \geq \tau_\text{T}$ as candidates. Then, it verifies whether $\text{sim}_\text{R} \geq \tau_\text{R}$. The drawback of the method is that it may generate too many candidates, leading to low search performance.

**Spatial-first method.** The method first finds the objects with $\text{sim}_\text{R} \geq \tau_\text{R}$ as candidates, and then filters the candidates whose $\text{sim}_\text{T} < \tau_\text{T}$. The method may also generate too many candidates, and cannot find similar objects fast.

**Spatial keyword search based method.** We can extend the IR-tree method [7] to support our spatio-textual similarity search as follows. Specifically, we construct an IR-tree to index all objects, where each node contains an MBR and an inverted file which maps a token to the child nodes containing the token. In particular, we store the spatio-textual objects in leaf nodes. Given a query $q = (R,T)$, the IR-tree based algorithm traverses the tree from the root to its leaf nodes. The algorithm takes an intermediate node $n$ as a candidate and visits its descendants, if 1) the spatial overlap $|q.R \cap n.R| \geq c_\text{R}$ and 2) the textual overlap $\sum_{t \in q.T \cap n.T} w(t) \geq c_\text{T}$, where $c_\text{R}$ and $c_\text{T}$ are thresholds derived from $\tau_\text{R}$ and $\tau_\text{T}$, which will be discussed respectively in Sections 3 and 4. For a leaf node corresponding to an object, we verify whether its spatial and textual similarities to $q$ are respectively not smaller than $\tau_\text{R}$ and $\tau_\text{T}$.

The algorithm may visit too many unnecessary nodes and lead to low search efficiency. We explain it by taking the objects in Figure 1 as an example. Using a maximum fan-out 3, we can construct an IR-tree as shown in Figure 2, where leaf nodes correspond to objects and intermediate nodes (i.e., $R_8$, $R_9$ and $R_{10}$) are MBRs bounding the objects. For simplicity, we use $R$ to represent both tree nodes and

826

their regions. Given query $q$ in Figure 1, the algorithm has to visit nodes $R_8$, $R_9$ and $R_{10}$ and verifies their leaf nodes to report the answer $\{o_2\}$. Obviously, it is unnecessary to visit the subtrees rooted at $R_9$ and $R_{10}$ as none of the five leaf nodes is similar to $q$. Therefore, the IR-`tree`-based methods have poor filtering power due to the hierarchical structure of the IR-`tree`. Moreover, the IR-`tree` maintains a tree based inverted index in each node for mapping tokens to its child nodes. Let $\mathcal{H}$ denote the height of the IR-`tree`. In the worst case, each token of every object needs to be indexed $\mathcal{H}$ times, and this results in high space complexity. To address these problems, in this paper we propose a novel method SEAL.

## 3. THE SEAL METHOD

In this section, we first introduce a filter-and-verification framework in Section 3.1, and then present a textual-based filtering algorithm in Section 3.2.

### 3.1 A Filter-and-Verification Framework

To answer a spatio-textual similarity search query efficiently, we want to prune dissimilar objects and only visit a small amount of objects that may be similar to the query. To this end, we propose a filter-and-verification framework.

**Step 1 - Filter:** We prune a large amount of objects which cannot be similar to query $q$, and find a *candidate* set $\mathcal{C}$, which is a superset of the answer set $\mathcal{A}$.

**Step 2 - Verification:** We verify the candidates generated in the filter step by checking whether spatial and textual similarities of each candidate are respectively not smaller than thresholds $\tau_R$ and $\tau_T$, and return the answer set $\mathcal{A}$.

In this paper, we focus on the filter step and propose efficient signature-based filtering algorithms. Consider a spatio-textual object $o \in \mathcal{O}$ and a query $q$. We denote their signatures as $S(o)$ and $S(q)$, where each $s \in S(o)$ (or $S(q)$) is called a *signature element* (or *element* for simplicity). A signature method must satisfy the following property:

*$o$ is similar to $q$ only if $S(o)$ and $S(q)$ are similar.*

Specifically, given a signature similarity function $\texttt{sim}(\cdot)$ and a threshold $c$ deduced from thresholds $\tau_R$ and $\tau_T$, the object $o$ is similar to $q$ only if $\texttt{sim}\big(S(q), S(o)\big) \geq c$.

A naïve method for generating candidates enumerates the signature of each object $o$ and checks if $\texttt{sim}(S(q), S(o)) \geq c$. If so, we add $o$ into candidate set $\mathcal{C}$. However this method is rather expensive and we want to build indexes to support efficient filtering, which will be discussed later.

Next we introduce our algorithm SEALSIG and Figure 3 illustrates the pseudo-code. We first generate the signatures for objects in $\mathcal{O}$ and build an index on top of the signatures (line 2). Then for a query $q$, we use the index to find its candidate set $\mathcal{C}$ (line 3). Finally, we verify the candidates in $\mathcal{C}$ and return the answer set $\mathcal{A}$ (line 4).

In this paper we focus on generating effective signatures, building indexes and devising efficient filtering algorithms.

### 3.2 A Textual-Based Filtering Algorithm

We introduce a textual-based filtering algorithm.

**Textual Signatures.** Since object $o$ similar to query $q$ must satisfy $\texttt{sim}_T(q,o) \geq \tau_T$, we can use the tokens in $o$ as its *textual signature*, i.e., $S_T(o) = o.T$, where each token $t \in o.T$ is a signature element. Similarly, we can also generate a textual signature for $q$, i.e., $S_T(q) = q.T$.

---

**Algorithm 1**: SEALSIG $(\mathcal{O}, q)$

**Input**: $\mathcal{O}$: An object set; $q$: A query
**Output**: $\mathcal{A}$: Answers of $q$

1 **begin**
2    Generate signatures for $\mathcal{O}$ and build index $\mathcal{I}$;
3    $\mathcal{C}$ = SIG-FILTER $(q, \mathcal{I})$ ;
4    $\mathcal{A}$ = SIG-VERIFY $(q, \mathcal{C})$ ;
5 **end**

---

**Function** SIG-FILTER$(q, \mathcal{I})$

**Input**: $q$: A query; $\mathcal{I}$: An inverted index
**Output**: $\mathcal{C}$: Candidate objects

1 **begin**
2    Initialize a candidate set $\mathcal{C} \leftarrow \emptyset$ ;
3    Generate signature, $S(q) \leftarrow$ GENSIG $(q)$ ;
4    Compute signature similarity threshold $c$ ;
5    **for** *each element $s$ in $S(q)$* **do**
6      Obtain objects in inverted list $\mathcal{I}(s)$ ;
7      Merge objects with $\texttt{sim}\big(S(q), S(o)\big) \geq c$ to $\mathcal{C}$ ;
8 **end**

---

**Function** SIG-VERIFY$(q, \mathcal{C})$

**Input**: $q$: A query; $\mathcal{C}$: A set of candidate objects
**Output**: $\mathcal{A}$: Answers of $q$

1 **begin**
2    **for** *each object $o \in \mathcal{C}$* **do**
3      **if** $\texttt{sim}_R(q,o) \geq \tau_R$ & $\texttt{sim}_T(q,o) \geq \tau_T$ **then**
         $\mathcal{A} \leftarrow \mathcal{A} \bigcup \{o\}$ ;
4 **end**

**Figure 3: A Filter-and-Verification Framework**

---

Then, we define the similarity between signatures $S_T(o)$ and $S_T(q)$ as the weight summation of their common tokens, i.e., $\texttt{sim}\big(S_T(q), S_T(o)\big) = \sum_{t \in S_T(q) \cap S_T(o)} w(t)$, and threshold $c_T = \tau_T \cdot \sum_{s \in S_T(q)} w(t)$. It is easy to prove that $\texttt{sim}_T(q,o) \geq \tau_T$ only if $\sum_{t \in S_T(q) \cap S_T(o)} w(t) \geq c_T$. For example, we can respectively generate textual signatures for $o_1$ and $q$ as $S_T(o_1) = \{t_1, t_2\}$ and $S_T(q) = \{t_1, t_2, t_3\}$. Given $\tau_T = 0.3$, the threshold $c_T$ can be computed as 0.57. Obviously, since $\texttt{sim}_T(q,o) \geq \tau_T$, we have $\sum_{t \in S_T(q) \cap S_T(o_7)} w(t) \geq c_T$.

**Indexing Structures.** To avoid enumerating every object $o \in \mathcal{O}$ for computing $\texttt{sim}\big(S(q), S(o)\big)$, we build an inverted index on top of the signatures. Formally, an inverted index $\mathcal{I}$ consists of a set of *inverted lists*, each of which maps an element $s$ to the objects containing the element, denoted by $\mathcal{I}(s)$. Figure 4 provides the inverted index of the objects in Figure 1. For example, since the signatures of object $o_3$ and $o_6$ contain element $t_4$, the inverted list of $t_4$ is $\{o_3, o_6\}$. In addition, since $o_2$ has textual signature $S_T(o_2) = \{t_1, t_3, t_2\}$, the object is contained in the inverted lists of $t_1$, $t_3$ and $t_2$.

**Filtering Algorithm.** Given a query $q$ with thresholds $\tau_R$ and $\tau_T$, the filtering algorithm SIG-FILTER in Figure 3 is utilized to filter dissimilar objects and find the candidates with signatures similar to $S(q)$, i.e., $\mathcal{C} = \{o \mid o \in \mathcal{O}, \texttt{sim}\big(S(q), S(o)\big) \geq c\}$. Specifically, the algorithm generates a signature for $q$, i.e., $S(q) \leftarrow \text{GENSIG}(q)$, and computes threshold $c$. For each element $s \in S(q)$, SIG-FILTER probes inverted list $\mathcal{I}(s)$ from the inverted index $\mathcal{I}$, and merges the objects in $\mathcal{I}(s)$ satisfying $\texttt{sim}\big(S(q), S(o)\big) \geq c$ to $\mathcal{C}$.

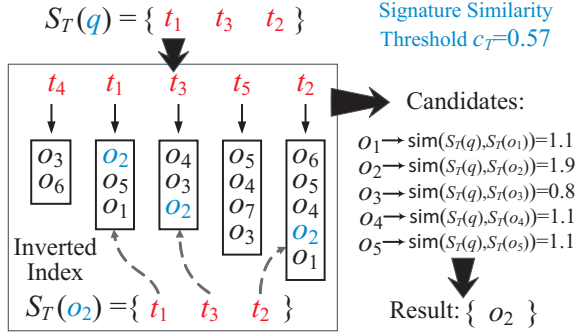

Figure 4: Textual signature based method.

EXAMPLE 2. *We use an example to illustrate how algorithm* SEALSIG *works using textual signatures, as shown in Figure 4. Consider the objects and query q with thresholds $\tau_R = 0.25$ and $\tau_T = 0.3$ in Figure 1. The algorithm generates the textual signature $S_T(o)$ for every object $o \in \mathcal{O}$, and then builds the inverted index on top of the signatures. Given query q, the algorithm* SIG-FILTER *generates its textual signature $S_T(q) = \{t_1, t_2, t_3\}$ and computes the threshold $c_T = 0.57$. Then, it probes the inverted lists of $t_1$, $t_3$ and $t_2$, and finds the candidate objects satisfying $\text{sim}(S(q), S(o)) \geq c_T$, i.e., $\mathcal{C} = \{o_1, o_2, o_3, o_4, o_5\}$. Finally,* SIG-VERIFY *verifies the candidates and reports the answer $\mathcal{A} = \{o_2\}$.*

The algorithm SEALSIG using textual signatures has a limitation that it fails to consider the spatial information. Recall that an object $o$ is similar to query $q$ if and only if textual similarity $\text{sim}_T(q, o) \geq \tau_T$ and spatial similarity $\text{sim}_R(q, o) \geq \tau_R$. Obviously, it is rather limited to only consider the textual information. For example, consider object $o_4$ in Figure 1. Although the textual similarity is larger than $\tau_T$, $o_4$ is not similar to $q$ since its region $R_4$ is dissimilar to $R_q$. To address this problem, we propose to generate spatial signatures for objects and queries, and devise efficient filtering algorithms using spatial signatures for *spatial pruning* in Section 4. In order to utilize spatial and textual pruning simultaneously, we develop more efficient *hybrid* filtering algorithms in Section 5.

## 4. GRID-BASED FILTERING ALGORITHM

In this section, we propose a *grid*-based filtering algorithm. We first define the grid-based signature in Section 4.1, and then devise a threshold-aware pruning technique in Section 4.2. Finally, we discuss how to select grid granularity to achieve better performance in Section 4.3.

### 4.1 Grid-Based Signatures

Different from textual signatures, as the spatial information has no inherent elements (e.g., tokens), it is not straightforward to generate spatial signatures for objects in $\mathcal{O}$ and query $q$. To address this challenge, we propose to partition the entire space $\mathcal{R}$ which is the MBR of the regions of all objects in $\mathcal{O}$, and generate a set of *grids*. Then, for an object $o$, we use the grids intersecting with region $o.R$ as its spatial signature. More formally, let $G = \{g_1, g_2, \ldots, g_{|G|}\}$ be a set of grids in space $\mathcal{R}$, where each grid is also an MBR, and the grids have the following properties.

1) *Completeness*: All grids cover the space, i.e., $\bigcup_{g \in G} = \mathcal{R}$.

2) *Disjointness*: Each pair of different grids is disjoinable, i.e., $\forall i, j$, if $i \neq j$, $g_i \cap g_j = \emptyset$.

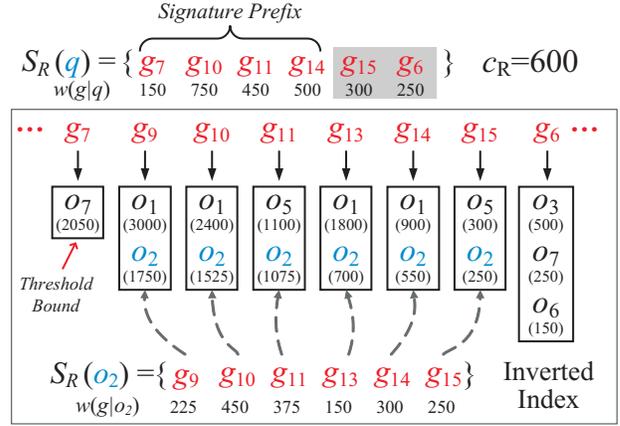

Figure 5: Threshold-aware pruning.

We only consider the *uniform* grids with equal size, i.e., $|g_i| = |g_j|$ ($i \neq j$) in this section, and will extend our techniques to support grids with different sizes in Section 5. Figure 1 provides an example of uniform grids $G = \{g_1, \ldots, g_{16}\}$ obtained by a $4 \times 4$ partition of the space $\mathcal{R}$ for regions $\{R_1, R_2, \ldots, R_7\}$. Then, we define the grid-based signature of object $o$ (query $q$), denoted by $S_R(o)$ ($S_R(q)$), as the grids intersecting with region $o.R$ ($q.R$) in Definition 4.

DEFINITION 4 (GRID-BASED SIGNATURE). *Given object $o$, the grid-based signature of $o$ is the grids in $G$ intersecting with region $o.R$, i.e., $S_R(o) = \{g \mid g \in G, g \cap o.R \neq \emptyset\}$.*

For example, Figure 5 shows the grid-based signature of object $o_2$, $S_R(o_2) = \{g_9, g_{10}, g_{11}, g_{13}, g_{14}, g_{15}\}$.

Next, we define similarity between grid-based signatures $S_R(q)$ and $S_R(o)$ as weight summation of their common grids, i.e., $\text{sim}(S_R(q), S_R(o)) = \sum_{g \in S_R(q) \cap S_R(o)} w(g \mid q, o)$, where $w(g \mid q, o)$ is the weight of grid $g$ with respect to query $q$ and object $o$. Intuitively, the weight captures the degree of spatial similarity between $q$ and $o$ contributed by $g$.

It is not straightforward to define grid weight due to the following reason. Ideally, if grid $g$ is shared by $q$ and $o$, the weight should be the area of the intersecting region of $q$ and $o$ in the grid, i.e., $w(g \mid q, o) = |q.R \cap o.R \cap g|$. However, in practice, it is expensive to compute $|q.R \cap o.R \cap g|$ at the query time. Thus, we propose to employ an upper bound of $|q.R \cap o.R \cap g|$ to estimate the weight as,

$$w(g \mid q, o) = \min\{w(g \mid q), w(g \mid o)\}, \quad (1)$$

where $w(g \mid o)$ ($w(g \mid q)$) is the weight of grid $g$ with respect to object $o$ (or query $q$), and can be estimated by $w(g \mid o) = |g \cap o.R|$ or $w(g \mid q) = |g \cap q.R|$.

In addition, we define the threshold $c_R$ to be the area of region $q.R$ multiplied by threshold $\tau_R$, i.e., $\tau_R \cdot |q.R|$. Now, we can prove that the grid-based signature satisfies the key property of signatures, as shown in Lemma 1.

LEMMA 1. *Spatial similarity satisfies $\text{sim}_R(q, o) \geq \tau_R$, only if $\text{sim}(S_R(q), S_R(o)) \geq c_R$.*

PROOF. The proofs are in our technical report [8]. □

For example, in Figure 5, based on the grids $\{g_1, \ldots, g_{16}\}$, we generate grid-based signatures, $S_R(q)$ and $S_R(o_2)$, for query $q$ and object $o_2$. We also compute grid weights, e.g., $w(g_{10} \mid q) = 750$, $w(g_{10} \mid o_2) = 450$. Thus, we obtain signature similarity $\text{sim}(S_R(q), S_R(o_2)) = 1375$. Obviously, we have $\text{sim}(S_R(q), S_R(o)) \geq \tau_R \cdot |q.R| = 600$.

## 4.2 Threshold-Aware Pruning

Based on grid-based signatures, a straightforward method to filter dissimilar objects is to simply employ the algorithm SIG-FILTER in Figure 3. The algorithm probes inverted lists of grids in signature $S_R(q)$, and inserts the objects satisfying $\text{sim}(S_R(q), S_R(o)) \geq c_R$ into candidate set $\mathcal{C}$. However algorithm SIG-FILTER has the following limitations. Firstly, for a large query region $q.R$, since it may generate many grids as signature for $q$, the algorithm needs to probe many inverted lists, which may be very expensive. Secondly, when probing the inverted list of a grid, we have to retrieve all objects in the list, resulting in high probing costs for long inverted lists.

To address the above-mentioned limitations, we devise a threshold-aware pruning technique in this section. The objective of our technique is two-fold: 1) to reduce the number of probed inverted lists; and 2) to reduce the number of objects retrieved from a probed inverted list. To this end, we employ the *prefix-filtering* [5] mentioned in Section 2.2.

To use prefix filtering, we first fix a *global order* on the generated signature elements of the objects in $\mathcal{O}$. Then, we sort the elements of every object based on the global order. In the filter step, for each object $o$, instead of using signature $S(o)$, we only select a *prefix* of the signature, denoted by $S^p(o)$. Similarly, we also select a signature prefix for the query $S^p(q)$. The signature prefixes must satisfy that $\sum_{s \in S(q) \cap S(o)} w(s) < c$ if $S^p(q) \cap S^p(o) = \emptyset$.

**Prefix Selection.** Given a global order of signature elements, consider signature $S(o) = \{s_1, s_2, \ldots, s_{|S(o)|}\}$ of object $o$, where $s_i$ is the $i$-th element based on the global order. To select prefix $S^p(o) = \{s_1, \ldots, s_p\}$, we can remove the last elements with weight summation smaller than $c$ and select the remaining elements as the prefix, as shown in Lemma 2.

LEMMA 2. *Given a similarity threshold $c$, the signature prefix $S^p(o) = \{s_1, \ldots, s_p\}$ can be selected as*

$$p = \min\{i\}, \text{ s.t. } \sum_{j=i+1}^{|S(o)|} w(s) < c. \quad (2)$$

**Grid Order.** In this paper, we sort the grids in ascending order of the number of the object regions intersecting with them[4]. Formally, let $\text{count}(g)$ denote the number of object regions in $\mathcal{O}$ intersecting with grid $g$, i.e., $\text{count}(g) = |\{R \mid R \cap g \neq \emptyset\}|$. Then, we sort all grids in $G$ in ascending order of $\text{count}(g)$. For example, in Figure 5 we sort the grids intersecting with $q.R$ as $\{g_7, g_{10}, g_{11}, g_{14}, g_{15}, g_6\}$. Observed from this figure, we have $c_R = 600$, $w(g_{15} \mid q) = 300$ and $w(g_6 \mid q) = 250$. According to Lemma 2, we can select signature prefix as $\{g_7, g_{10}, g_{11}, g_{14}\}$, i.e., $p = 4$, because any shorter prefix $(p < 4)$ may cause the reduction of weights larger than threshold $c_R$, i.e., $\sum_{j=p+1}^{|S(o)|} w(g \mid q, o) \geq c_R$.

Therefore, instead of considering all elements in $S(q)$, we only need to probe the inverted lists of the ones in $S^p(q)$, which can reduce the filtering complexity.

**Inverted Index with Threshold Bounds.** To further improve the performance, we propose to reduce the number of retrieved objects in a probed inverted list. Recall that, for an object $o$, we only need to consider the signature elements

---
[4]Note that the global grid order will influence the performance of filtering. We do not study the problem in this paper due to the space limitation, and take it as future work.

---

**Function** SIG-FILTER$^+$($q$, $\mathcal{I}$)

**Input**: $q$: A query; $\mathcal{I}$: An inverted index
**Output**: $\mathcal{C}$: Candidate objects
1 **begin**
2    Initialize candidate set $\mathcal{C} \leftarrow \emptyset$ ;
3    Generate signature, $S(q) \leftarrow$ GENSIG $(q)$ ;
4    Compute signature similarity threshold $c$ ;
5    Select prefix $S^p(q)$ for query $q$ ;
6    **for** *each element $s$ in $S^p(q)$* **do**
7      Find the objects $\mathcal{I}^c(s)$ from inverted list $\mathcal{I}(s)$ ;
8      $\mathcal{C} \leftarrow \mathcal{C} \bigcup \mathcal{I}^c(s)$ ;
9 **end**

**Figure 6: Filtering with threshold-aware pruning**

in prefix $S^p(o)$ rather than $S(o)$. Thus, when probing the inverted list $\mathcal{I}(s)$ of element $s \in S(o)$, we only need to retrieve the objects in $\mathcal{I}(s)$ containing $s$ in their signature prefixes given threshold $c$, i.e., $\mathcal{I}^c(s) = \{o \mid o \in \mathcal{I}(s), s \in S^p(o)\}$.

A challenging problem is to efficiently compute $\mathcal{I}^c(s)$ based on various thresholds $c$ for different queries. To address this problem, we augment a *threshold bound* to each object in every inverted list. Specifically, for an object $o$ in an inverted list $\mathcal{I}(s)$ of element $s$, we maintain a threshold upper bound $\bar{c}_s(o)$, which represents the maximum threshold that we keep for $s$ in $o$'s signature prefix. Thus, if threshold $c > \bar{c}_s(o)$, object $o$ can be pruned from $\mathcal{I}^c(s)$, as shown below.

LEMMA 3. *Let $s_i$ be the $i$-th signature element in $S(o)$ (where $1 \leq i \leq |S(o)|$) and $c$ be a signature similarity threshold. The object $o$ can be pruned from $\mathcal{I}^c(s_i)$ if*

$$c > \bar{c}_{s_i}(o) = \sum_{j=i}^{|S(o)|} w(s_j). \quad (3)$$

We store bound $\bar{c}_s(o)$ for each object $o$ in inverted list $\mathcal{I}(s)$, and sort the objects in descending order of the bounds. Thus, given a threshold $c$, we can efficiently find $\mathcal{I}^c(s) = \{o \mid o \in \mathcal{I}(s), \bar{c}_s(o) \geq c\}$, when probing the inverted list of element $s$. Figure 5 provides the inverted lists of eight grids. For example, the inverted list of grid $g_{14}$ contains objects $\{o_1, o_2\}$, each of which is associated with a threshold bound, e.g., $\bar{c}_{g_{14}}(o_1) = 900$. Given threshold $c_R = 600$, we only retrieve $o_1$ when probing the inverted list of $g_{14}$, because bound $\bar{c}_{g_{14}}(o_2) = 550 < c_R$.

**Threshold-Aware Pruning.** Based on the technique mentioned above, we devise an improved filtering algorithm SIG-FILTER$^+$ in Figure 6. Compared with algorithm SIG-FILTER, SIG-FILTER$^+$ only selects signature prefix $S^p(q)$ for query $q$. For each element $s \in S^p(q)$, it only retrieves objects in $\mathcal{I}^c(s)$ instead of $\mathcal{I}(s)$, and merges the objects to the candidates $\mathcal{C}$. We use the following example to illustrate how algorithm SIG-FILTER$^+$ works using grid-based signatures.

EXAMPLE 3. *Consider the objects $\mathcal{O}$ and query $q$ with thresholds $\tau_R = 0.25$ and $\tau_T = 0.3$ in Figure 1. We generate grid-based signatures and build the inverted index with threshold bounds as shown in Figure 5. Given query $q$, we first generate its signature prefix, $S_R^p(q) = \{g_7, g_{10}, g_{11}, g_{14}\}$ based on threshold $c_R = 600$ according to Lemma 2. Then, for each element in $S_R(q)$, we probe its inverted list and only retrieve the objects with bounds no smaller than $c_R$, and obtain the candidates $\mathcal{C}_R = \{o_1, o_2, o_5, o_7\}$. Finally, algorithm SIG-VERIFY reports the answer of $q$, i.e., $\mathcal{A} = \{o_2\}$.*

Algorithm SIG-FILTER$^+$ can be also applied to textual signatures. Specifically, we can sort tokens in descending order of their idfs, and build the inverted index with threshold bounds. Given query $q$, we only consider the tokens in the signature prefix, denoted by $S_T^p(q)$, and only retrieve the objects with thresholds no smaller than $c_T$ in each inverted list. For example, we only retrieve inverted lists of $t_1$ and $t_3$ in Figure 4, and obtain candidates $C_T = \{o_1, o_2, o_3, o_4, o_5\}$. Notice that algorithm SIG-FILTER$^+$ may produce different sizes of candidates when using different signatures.

### 4.3 Grid Granularity Selection

An essential task in algorithm SIG-FILTER$^+$ is to generate grid-based signatures, i.e, GENSIG. Obviously, the performance of SIG-FILTER$^+$ is affected by grid granularity and a key challenge is to select an *appropriate* grid granularity. More specifically, coarse granularity achieves high filtering performance but weakens the filtering power and leads to low verification performance. On the contrary, the fine granularity reduces the number of candidates, but leads to low filtering performance. For instance, in Example 3, $o_5$ is a candidate of query $q$ as $o_5.R$ and $q.R$ share grid $g_{15}$. However, the two regions do not intersect with each other at all, and thus $o_5$ can be actually pruned.

To alleviate the problem, we propose a method for selecting grid granularity in this section. We introduce a probabilistic model to measure the *expected query cost* for grids of specific granularity. To answer a query $q$, the overall cost $\text{cost}(q)$ consists of filtering cost $\text{cost}_F(q)$ and verification cost $\text{cost}_V(q)$, i.e., $\text{cost}(q) = \text{cost}_F(q) + \text{cost}_V(q)$. The filtering cost $\text{cost}_F(q)$ depends on the number of objects retrieved from the inverted index, i.e., $\text{cost}_F(q) = \pi_1 \cdot \sum_{g \in S_R^p(q)} |\mathcal{I}^c(g)|$, where $\pi_1$ is the average cost of retrieving an object from an inverted list and merging it to candidates. On the other hand, verification cost $\text{cost}_V(q)$ depends on the number of candidates, i.e., $\text{cost}_V(q) = \pi_2 \cdot |C|$, where $\pi_2$ is the average cost of verifying an object. Thus, we have $\text{cost}(q) = \pi_1 \cdot \sum_{g \in S_R^p(q)} |\mathcal{I}^c(g)| + \pi_2 \cdot |C|$. For example, we have $\text{cost}(q) = 6\pi_1 + 4\pi_2$ for query $q$ in Figure 5.

Notice that the above analysis is based on a single query. To analyze the expected query cost of grid set $G$ with specific granularity, we suppose that we have a query workload $Q$. Then, each grid $g \in G$ has a probability $P(g)$ representing the likelihood that $g$ is used by queries. In addition, when inverted list $\mathcal{I}(g)$ is probed by different queries, the returned objects $\mathcal{I}^c(g)$ may be different. For ease of analysis, we consider the worst case that all objects need to be returned, i.e., $|\mathcal{I}^c(g)| = |\mathcal{I}(g)|$. Thus, we can estimate the expected query cost of all grids in $G$ with the specific granularity as

$$\widehat{\text{cost}}(G) = \pi_1 \cdot \sum_{g \in G} P(g) \cdot |\mathcal{I}(g)| + \pi_2 \cdot \overline{|C|}, \quad (4)$$

where $\overline{|C|}$ is the average size of candidates given the query workload. Now, we can define the grid granularity selection problem: Find the best set of grids $G$ with specific granularity that minimize the expected cost $\widehat{\text{cost}}(G)$.

Since it is intractable to solve the problem by considering arbitrary grid partition schemes, we devise an approximate algorithm for selecting grid granularity as illustrated in Figure 7. The basic idea is to decompose the underlying space $\mathcal{R}$ into a *grid tree* with height $\mathcal{H}$, where the grids in level $l$, denoted by $G^l$, is obtain by a $2^l \times 2^l$ partition of space $\mathcal{R}$. For example, at level 0, there is 1*1 grid, level 1 partitions $\mathcal{R}$ into $2*2$ grids, and so forth. Therefore, we reduce grid granularity selection to the problem of finding the best level $l^*$ in the grid tree to minimize the expected cost.

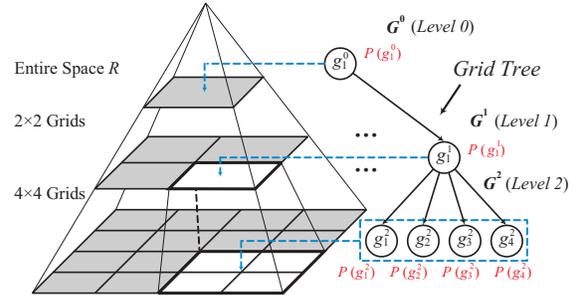

**Figure 7: Grid Granularity Selection**

We devise an approximate algorithm to solve this problem. We traverse the grid tree from the root to its leaves, and compute the expected cost of each level. For two adjacent levels, $l$ and $l+1$ ($0 \leq l < \mathcal{H}$), we compute the benefit of the partitioning as $\mathcal{B}(l, l+1) = \widehat{\text{cost}}(G^l) - \widehat{\text{cost}}(G^{l+1})$. If the benefit is smaller than a threshold $\overline{B}$, the algorithm terminates, where $\overline{B} > 0$ is used to balance efficiency and storage. Then, we prove that for any $\overline{B}$, we can find a level $\bar{l}$, which satisfies that for any level $l > \bar{l}$ the filtering benefit $\mathcal{B}(l, l+1) < \overline{B}$ as formalized in Lemma 4.

LEMMA 4. $\forall \overline{B} > 0$, there exists a level $\bar{l}$ which satisfies that for levels $l > \bar{l}$ filtering benefit $\mathcal{B}_F(l, l+1) < \overline{B}$.

We briefly show the correctness of Lemma 4 (The proof is in [8]). Based on Equation (4), we have $\mathcal{B}_F = \sum_{g^l \in G^l} \mathcal{B}_F(g^l)$, where $\mathcal{B}_F(g^l)$ is the benefit of partitioning grid $g^l$ into fine-grained grids. We use the example of partitioning $g_1^1$ into $\{g_1^2, g_2^2, g_3^2, g_4^2\}$ to show how to compute $\mathcal{B}_F(g^l)$. Based on the probabilistic theory, we have $P(g_1^1) = \sum_i P(g_i^2) - \sum_{i \neq j} P(g_i^2 g_j^2) + \sum_{i,j,k} P(g_i^2 g_j^2 g_k^2) - P(g_1^2 g_2^2 g_3^2 g_4^2)$. Since no query region can intersects with three grids, $P(g_i^2 g_j^2 g_k^2) = 0$. We denote $\sum_{i \neq j} P(g_i^2 g_j^2) + P(g_1^2 g_2^2 g_3^2 g_4^2)$ as $\theta$. Thus, the benefit of partitioning $g_1^1$ is $\pi_1 \cdot \left( \sum_{g_i^2} P(g_i^2) \cdot (|\mathcal{I}(g_1^1)| - |\mathcal{I}(g_i^2)|) - \theta \cdot |\mathcal{I}(g_1^1)| \right)$. With the increase of $l$, the benefit of partitioning a grid becomes less and less significant, since 1) $|\mathcal{I}(g_1^l)| - |\mathcal{I}(g_i^{l+1})|$ eventually decreases, and 2) $\theta$ eventually increases since it becomes more likely that queries intersect with more grids. Thus, we can find a level $\bar{l}$ which satisfies that for levels $l > \bar{l}$ filtering benefit $\mathcal{B}_F(l, l+1) < \overline{B}$.

Verification benefit $\mathcal{B}_V$ also has the similar property. However, it is difficult to analyze $\mathcal{B}_V$, as estimating the average candidate size $\overline{|C|}$ is very hard. Thus, we take theoretical analysis as a future work and only show the experimental results in Section 6.

## 5. HYBRID FILTERING ALGORITHMS

In this section, we develop hybrid filtering algorithms to simultaneously utilize textual and spatial signatures. We first introduce the hash-based hybrid signature and present a filtering algorithm based on the signature in Section 5.1, and then propose the hierarchical hybrid signature to further improve the performance in Section 5.2.

### 5.1 Hash-Based Hybrid Signature

A straightforward method is to respectively apply algorithm SIG-FILTER$^+$ using textual and grid-based signatures, and compute the intersection of candidate sets $C_T$ and $C_R$.

This method, although reducing the number of candidates, may increase the number of objects retrieved from inverted indexes, and thus result in higher filtering cost.

In order to reduce the filtering cost, we propose the hash-based hybrid signature to efficiently find possibly similar objects. The basic idea is to hash tokens and grids of each *same* object into buckets using a hashing function, and then use each bucket as a hybrid signature element.

DEFINITION 5 (THE HASH-BASED HYBRID SIGNATURE). *For each object $o$ with textual signature $S_T(o)$ and grid-based signature $S_R(o)$, the hash-based hybrid signature of $o$ is defined by $S_H(o) = \{h = (t,g) \mid t \in S_T(o), g \in S_R(o)\}$, where $h$ is a hash value by hashing $t$ and $g$ into a bucket.*

Based on hash-based hybrid signatures, we develop a more efficient filtering algorithm HYBRID-SIG-FILTER$^+$ in Figure 8. The algorithm respectively generates textual and grid-based signatures $S_T(o)$ and $S_R(o)$ for each object $o \in \mathcal{O}$. It hashes tokens in $S_T(o)$ and grids in $S_R(o)$ into buckets to generate a hybrid signature $S_H(o)$. Then, the algorithm builds an inverted index $\mathcal{I}$ for the hybrid signatures generated from all objects in $\mathcal{O}$. Compared with the inverted list introduced in Section 4.2, we augment both spatial and textual threshold bounds for each object $o$ in each inverted list of element $h$, denoted by $\bar{c}_h^T(o)$ and $\bar{c}_h^R(o)$. The two bounds can be computed according to Lemma 3, and satisfy that if either $c_T > \bar{c}_h^T(o)$ or $c_R > \bar{c}_h^R(o)$, $o$ can be safely pruned from the inverted list of element $h$. Moreover, to avoid generating too many inverted lists, we introduce a constraint of index sizes to guarantee that the number of hash buckets is smaller than a given number, which is explained in Section 5.2.

Given query $q$, the algorithm respectively generates textual and grid-based signatures $S_T(q)$ and $S_R(q)$, and computes signature similarity thresholds $c_T$ and $c_R$. Then, it respectively selects prefixes $S_T^p(q)$ and $S_R^p(q)$. Next, for each token $t \in S_T^p(q)$, the algorithm examines each grid $g \in S_R^p(q)$, and computes the hash-based signature element $h = (t,g)$. Using element $h$, the algorithm probes its inverted list $\mathcal{I}(h)$ and retrieves the objects satisfying $\bar{c}_h^T \geq c_T$ and $\bar{c}_h^R \geq c_R$, denoted by $\mathcal{I}^{\{c_R,c_T\}}(h) = \{o \in \mathcal{I}(h) \mid \bar{c}_h^T(o) \geq c_T, \bar{c}_h^R(o) \geq c_R\}$. Finally, the algorithm merges the retrieved objects $\mathcal{I}^{\{c_R,c_T\}}(h)$ to the candidate set $\mathcal{C}$.

EXAMPLE 4. *Figure 9 shows how algorithm* HYBRID-SIG-FILTER$^+$ *works. For each object, the algorithm hashes its tokens and grids into buckets to generate a hybrid signature. For example, the signature of object $o_1$ consists of elements such as $(t_1, g_{10})$, $(t_1, g_{11})$, $(t_1, g_{14})$, etc. Given query $q$ with prefixes $S_T^p(q) = \{t_1, t_3\}$ and $S_R^p(q) = \{g_7, g_{10}, g_{11}, g_{14}\}$, the algorithm obtains hash-based hybrid signatures, and probes the corresponding inverted lists. When probing a list, the algorithm only retrieves the objects satisfying $\bar{c}_h^T(o) \geq c_T$ and $\bar{c}_h^R(o) \geq c_R$. For example, the inverted list of element $(t_1, g_{14})$ only returns $o_1$. Finally, the algorithm produces candidates $\mathcal{C} = \{o_1, o_2, o_5\}$.*

## 5.2 Hierarchical Hybrid Signatures

Algorithm HYBRID-SIG-FILTER$^+$ generates hybrid signatures using grids with fixed granularity. As different regions can use grids with different granularities, this method has the following limitations. Generating coarse-grained grids

**Function** HYBRID-SIG-FILTER$^+$($q$, $\mathcal{I}$)
  **Input**: $q$: A query; $\mathcal{I}$: A hybrid inverted index
  **Output**: $\mathcal{C}$: Candidate objects
1 **begin**
2   | Initialize candidate set $\mathcal{C} \leftarrow \emptyset$ ;
3   | Generate signatures $S_T(q)$ and $S_R(q)$ ;
4   | Compute signature similarity thresholds $c_T$ and $c_R$ ;
5   | Select prefixes $S_T^p(q)$ and $S_R^p(q)$ ;
6   | **for** *each token $t$ in $S_T^p(q)$* **do**
7   |   | **for** *each grid $g$ in $S_R^p(q)$* **do**
8   |   |   | Compute the hybrid signature $h = (t,g)$ ;
9   |   |   | Find object list $\mathcal{I}^{\{c_R,c_T\}}(h)$ from $\mathcal{I}$ ;
10  |   |   | $\mathcal{C} \leftarrow \mathcal{C} \bigcup \mathcal{I}^{\{c_R,c_T\}}(h)$ ;
11 **end**

**Figure 8: Hybrid signature based filtering algorithm**

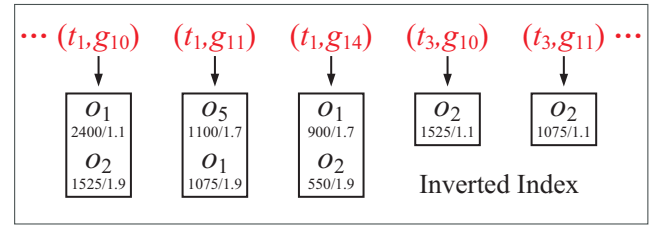

**Figure 9: Example of algorithm Hybrid-Sig-Filter$^+$**

for small regions may weaken the filtering power and introduce more candidates. For example, the element $(t_1, g_{11})$ in Figure 9 involves a dissimilar object $o_5$ as the estimated grid weight $w(g \mid o_5) = 200$ is much larger than the real weight $w(g \mid o_5, q) = 0$. On the other hand, generating fine-grained grids for large regions may involve too many useless signature elements, leading to high costs of both storage of inverted lists and filtering. In Figure 1, any fine-grained grid covered by $g_{14}$ is useless to region $R_1$, because $g_{14}$ has already provided an accurate grid weight for $R_1$.

In order to address the problem, we propose to judiciously select hierarchical grids for each token $t$ given a constraint of index sizes (i.e., the maximum number of hybrid signature elements), and generate hierarchical hybrid signatures to improve the performance. We first formalize the hierarchical hybrid signature selection problem as follows.

**Hierarchical hybrid signature selection.** Intuitively, our objective is to select at most $m_t$ hierarchical grids for the objects containing each token $t$ and optimize the *filtering power*. As mentioned in Section 4.3, optimizing the filtering power can be reduced to minimizing filter and verification costs. Since verification is the bottleneck as shown in Section 6.3, we focus on minimizing verification cost, i.e., the average size of candidates $\overline{|\mathcal{C}|}$ in this section. It is known that the estimation of $\overline{|\mathcal{C}|}$ is very difficult, so we consider a simplified version of the problem.

Ideally, suppose that we have a set of grids with *finest* granularity, such that each finest grid is totally covered by or exclusive from object regions. Using the finest grids, we can obtain the most compact candidate set satisfying $|\mathcal{C}| = |\mathcal{A}|$. Obviously, the amount of the finest grids must be very huge. Therefore, given a number $m_t$, we need to merge the finest grids to at most $m_t$ hierarchical grids. Different grids merged from the finest grids may lead to different upper bounds of grid weights (See Section 4.1) and result in different candidate sizes. To measure quality of the grids, we introduce the *error* of each grid. Formally, we define the error as follows. Consider a grid $g$ with inverted list $\mathcal{I}(g)$. Using uniform assumption, we can estimate the expected size of inverted list as $\widehat{|\mathcal{I}(g)|} = \sum_{o \in \mathcal{I}(g)} \frac{|g \cap o.R|}{|g|}$, where $\frac{|g \cap o.R|}{|g|}$ is the probability that query region $q.R$ intersects with each object region $o.R$. For example, consider grid $g_{13}$ in Figure 1. We can compute $\widehat{|\mathcal{I}(g_{13})|} = (|g_{13} \cap R_1| + |g_{13} \cap R_2|)/|g| = 1.7$.

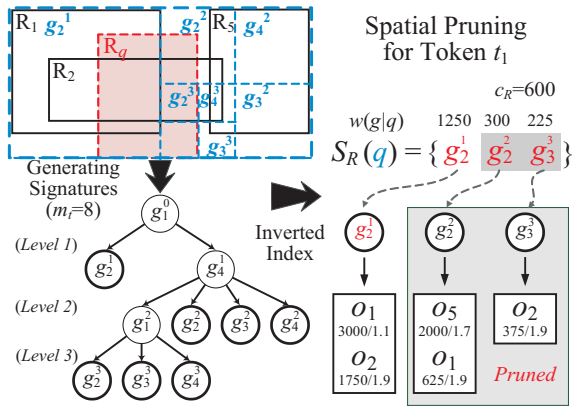

**Figure 10: Hierarchical hybrid signatures of $t_1$**

DEFINITION 6 (ERROR OF GRID). *Consider grid $g$ covering finest grids $\{g_1^f, \ldots, g_l^f\}$. The error of $g$, denoted by* ERROR$(g)$, *is* $\sum_{g_i^f} (\widehat{|\mathcal{I}(g)|} - \widehat{|\mathcal{I}(g_i^f)|})^2$.

Based on the errors of grids, we formally define our hierarchical hybrid signature selection (HSS) problem as follows.

DEFINITION 7 (THE HSS PROBLEM). *Given a token $t$ and a number $m_t > 1$, find at most $m_t$ hierarchical grids $G_t$ such that* $\sum_{g \in G_t}$ ERROR$(g)$ *is minimized.*

We can prove that the HSS problem is NP-hard by a reduction from a known NP-hard problem, a rectangular partitioning problem in [15].

THEOREM 1. *The HSS problem is NP-Hard.*

To solve the HSS problem, we propose a greedy algorithm to find the grids with the minimum errors, as shown in Figure 11. Given all objects indexed by a token $t$, i.e., $\mathcal{I}(t)$, the algorithm first constructs a grid tree, and initializes a priority queue where elements are sorted in descending order of their scores. Then, it inserts the root with its grid error as the score. Here, for a node $n$, the algorithm approximately computes its error based on the child nodes, ERROR$(n) = \sum_{\text{child}(n)} (\widehat{|\mathcal{I}(n)|} - |\mathcal{I}(\widehat{\text{child}}(n))|)^2$. Next, the algorithm traverses the grid tree until the queue is empty as follows. It removes the front element in the queue, denoted by $n$. If $n$ is a leaf node, the algorithm inserts it into $\mathcal{G}_t$. If $n$ is an intermediate node, the algorithm examines whether to split $n$ using its child nodes $\mathcal{N}_c$. If the number of grids after the splitting is larger than $m_t$, i.e., $|\mathcal{G}_t| + |\mathtt{Q}| + |\mathcal{N}_c| - 1 > m_t$, the algorithm does not split the node and only inserts $n$ into $\mathcal{G}_t$; otherwise, it inserts the child nodes of $n$ into the queue.

Figure 10 provides an example to show how algorithm HSS-Greedy generates hierarchical hybrid signatures for regions $\{R_1, R_2, R_5\}$ of token $t_1$. The algorithm first enqueues

---

**Algorithm 2**: HSS-Greedy $(\mathcal{I}(t), m_t)$

**Input**: $\mathcal{I}(t)$: objects containing token $t$; $m_t$: A number
**Output**: $\mathcal{G}_t$: The selected grids

1 **begin**
2     Construct a grid tree GT for objects in $\mathcal{I}(t)$ ;
3     Initialize a priority queue Q ;
4     Q.Enqueue $(\mathtt{GT}.root, \mathrm{ERROR}(\mathtt{GT}.root))$ ;
5     **while** Q *is not empty* **do**
6       $n \leftarrow$ Q.Dequeue() ;
7       **if** $n$ *is a leaf node* **then**   $\mathcal{G}_t \cup \{n\}$ ;
8       **else**
9          $\mathcal{N}_c \leftarrow$ Child nodes of $n$ ;
10          **if** $|\mathcal{G}_t| + |\mathtt{Q}| + |\mathcal{N}_c| - 1 > m_t$ **then**   $\mathcal{G}_t \cup \{n\}$ ;
11          **else**   **for** *Each child node $n_c$ in $\mathcal{N}_c$* **do**
12              Q.Enqueue $(n_c, \mathrm{ERROR}(n_c))$ ;

13 **end**

**Figure 11: Greedy Algorithm for the HSS problem.**

the root grid $g_1^0$, i.e., the space $\mathcal{R}$. Then, the algorithm repeatedly dequeues the front element, and enqueues its child nodes. For example, since ERROR$(g_4^1) = 0.1 >$ ERROR$(g_2^1) = 0.09$, the algorithm dequeues grid $g_4^1$ and enqueues its child nodes. Thus, given $m_t = 8$, we obtain hierarchical hybrid signatures represented as bold circles in Figure 10.

**Hierarchical signature-based filtering algorithm.** Using the generated hierarchical hybrid signatures, we can improve the performance of algorithm HYBRID-SIG-FILTER$^+$ (in Figure 8) as follows. For each token $t$, we first fix a global order of its hierarchical grids. We sort the grids in ascending order of their levels in the grid tree. For grids in the same level, we sort them in ascending order of the number of object regions intersecting with them. For example, based on the global order, the hierarchical hybrid signature of token $t_1$ can be sorted as $\{g_2^1, g_3^1, g_4^1, g_2^2, g_2^3, g_3^3, g_4^3\}$. Then, we can employ algorithm HYBRID-SIG-FILTER$^+$ to filter dissimilar objects, as illustrated in the following example.

EXAMPLE 5. *Figure 10 provides an example of algorithm* HYBRID-SIG-FILTER$^+$ *using hierarchical hybrid signatures. Consider token $t_1$ in textual signature prefix $S_T^p(q)$. We can generate three hierarchical grids $\{g_2^1, g_2^2, g_3^3\}$ and compute the weight $w(g \mid q)$ for each grid. Using threshold-aware pruning, we can prune inverted lists of grids $g_2^1$ and $g_2^2$, and only need to probe the inverted list of hybrid signature $(t_1, g_2^1)$. Similarly, we probe the inverted lists of other tokens in $S_T^p(q)$ and obtain a candidate set $\mathcal{C} = \{o_1, o_2\}$. Notice that the candidate set is more compact than the one in Example 4.*

## 6. EXPERIMENTS

In this section, we report experimental results. We extended the state-of-the-art spatial keyword search method IR-tree [7] to support spatio-textual similarity search as mentioned in Section 2.3. We compared our SEAL method with this method.

### 6.1 Experiment Setup

We used two datasets. The first one was a real dataset Twitter. We collected 60 million tweets from May 2011 to August 2011, among which about 13 million tweets had locations (i.e., points with longitudes and latitudes). We randomly selected 1 million users, and obtained ROI objects

832

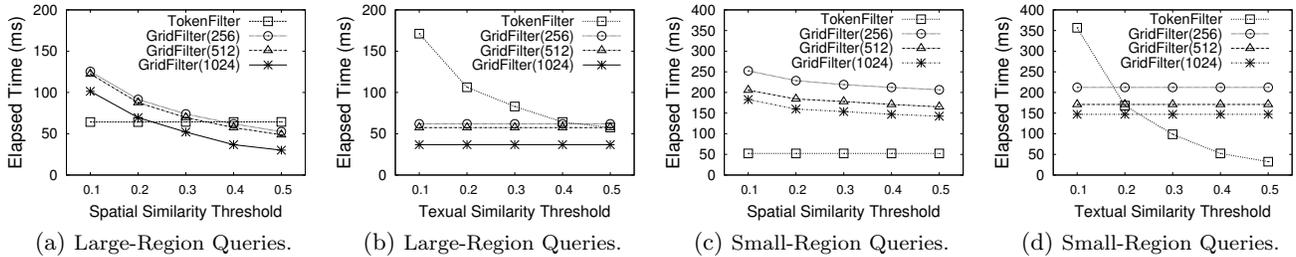
Figure 12: TokenFilter vs. GridFilter on the Twitter data set.

Table 1: Data statistics and index sizes.

|  | Twitter | USA |
|---|---|---|
| Object number | 1 million | 1 million |
| Avg region area (sq.km.) | 115 | 5 |
| Entire space (million sq.km.) | 1342 | 473 |
| Avg token number | 14.3 | 12.5 |
| Data size (GB) | 0.34 | 0.22 |
| IR-tree size (GB) | 2.37 | 1.14 |
| TokenInv size (GB) | 0.17 | 0.07 |
| GridInv (1024) size (GB) | 0.018 | 0.016 |
| HashInv (1024) size (GB) | 0.57 | 0.44 |
| HierarchicalInv size (GB) | 0.30 | 0.34 |

from their profiles. On the one hand, we selected *frequent* words in her tweets as the token set for each user, and the average token number of objects was 14.3. On the other hand, we took the MBR of all of her tweets as each user's active region, and the average area of the regions was 115 square kilometers (sq.km.). Specifically, the distributions of the region sizes were: 0.0001 sq.km. (4.4%), 0.01 sq.km. (15.4%), 1 sq.km. (29.7%), 100 sq.km. (73%), etc, where "$x$ sq.km. ($y$%)" means there are $y$% regions with areas not greater than $x$ sq.km. We can see that the regions had various sizes due to different spatial distributions of users' tweets, and most of regions sizes were not very large. Note that more complicated methods can be used to obtain the active regions from users' tweets. For example, we can compute multiple active regions for each user by clustering tweets' locations. We take it as a future work, and only focus on the general spatio-textual similarity search problem in this paper. In addition, we generated two query sets to evaluate the performance of different algorithms. Each query set contained 100 queries.

**Large-Region Queries**: We generated a set of queries with large regions. The average query token number was 6.97, and the average area of regions was 554 sq.km, which was equivalent to the area of a district.

**Small-Region Queries**: We generated a set of queries with small regions. The average query token number was 12.9, and the average area of query regions was 0.44 sq.km, which was equivalent to the area of a small neighborhood.

We also used a synthetic dataset by combing POIs in USA and the publication records in DBLP. We selected 1 million POIs from the USA dataset as centers and extended the POIs with random widths and heights to generate regions. Then, we randomly distributed publication records to the regions to generate token sets. The average area of regions was 5.4 sq.km and the average token number was 12.5. We also generated 100 large-region queries and 100 small-region queries for this dataset. Note that the experimental results on the two datasets were similar. Due to the space constraints, we only provide experimental results of method comparison on the USA data set in Section 6.5.

The IR-tree index was disk-resident, and its page size was 4KB. The inverted indexes of our signature-based methods were also disk-resident, and we maintained an index that mapped each signature element to the disk offset of its inverted list in memory. Note that this index was small enough to be maintained in memory. For example, for the Twitter dataset, the index only occupied 19 MB. Table 1 summarizes the data statistics and index sizes. For simplicity, we respectively use TokenInv, GridInv, HashInv and HierarchicalInv to represent inverted indexes of textual, grid-based, hash-based hybrid and hierarchical hybrid signatures. For GridInv and HashInv we also present the granularity. For example, GridInv (1024) represents the GridInv index with granularity $1024 \times 1024$. Due to space constraints, we only show sizes of the indexes we used in method comparison in Section 6.5. Besides, we varied both spatial and textual similarity thresholds from 0.1 to 0.5, and the default value of the thresholds was 0.4.

In the paper, we only show the running time and the numbers of candidate numbers of different methods are in our technical report [8].

All the programs were implemented in JAVA and all the experiments were run on the Ubuntu machine with an Intel Core 2 Quad X5450 3.00GHz processor and 4 GB memory.

### 6.2 Token Filter vs. Grid Filter

We first evaluated algorithm SIG-FILTER$^+$ in Figure 6 using textual signatures and grid-based signatures, which are respectively denoted by TokenFilter and GridFilter. We examined the GridFilter using grid-based signatures with different granularities. Figure 12 provides the experimental results. We can see that the performance of TokenFilter and GridFilter depended on the similarity thresholds, $\tau_R$ and $\tau_T$. Observed from Figure 12(a), TokenFilter outperformed GridFilter for small $\tau_R$. However, with the increase of $\tau_R$, GridFilter became faster. For example, given $\tau_R = 0.5$, GridFilter (1024) took 30 milliseconds, while TokenFilter took 64 millisecond. Similarly, with the increase of $\tau_T$, TokenFilter became better (See Figures 12(b) and 12(d)). Therefore, it is better to combine both filters instead of using either one individually.

### 6.3 Evaluating Different Grid Granularities

We then evaluated the performance of GridFilter with different grid granularities. We partitioned the entire space into $p \times p$ uniform grids as mentioned in Section 4, where $p$ denotes the granularity. Then, we ran GridFilter based on different granularities and compared the filtering time and verification time. Figure 13 shows the experimental results. We can see that with the increase of granularity, the verification time always decreased. For example, the verification time decreased from 466 milliseconds at granularity 64 to 90 milliseconds at granularity 8192 in Figure 13(b). Moreover, the decrease of verification cost became more and more insignificant as the granularity increased. The filtering step was not always improved with the increase of granularity, which was consistent with our cost-based analysis in Section 4.3. For example, in Figure 13(a), the filtering time first decreased and then increased after granularity 1024.

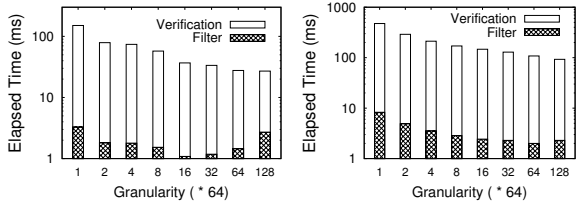

(a) Large-Region Queries.   (b) Small-Region Queries.
Figure 13: Evaluation on grid granularity selection on the Twitter data set.

### 6.4 Evaluating Hybrid Filtering Algorithms

We evaluated hybrid filtering algorithms mentioned in Section 5. We first compared the algorithm using hash-based hybrid signatures (HYBRIDFILTER) with GRIDFILTER. Figure 14 shows the experimental results, where G and H respectively stand for GRIDFILTER and HYBRIDFILTER. We can see that HYBRIDFILTERs significantly outperformed GRIDFILTERs for both large-region and small-region queries. For example, in Figure 14(a), HYBRIDFILTERs with different granularities (i.e., 256, 512, 1024) were one order of magnitude faster than GRIDFILTERs with the same granularity. This is because HYBRIDFILTER utilized both spatial and textual pruning simultaneously and had larger filtering power than GRIDFILTER with only spatial pruning.

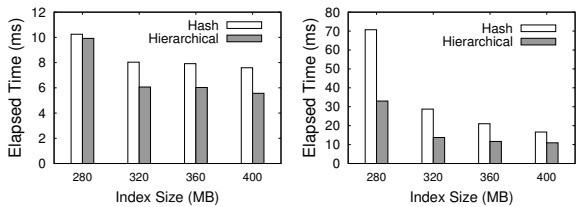

(a) Large-Region Query.   (b) Small-Region Query.
Figure 15: Comparison of hybrid signatures on the Twitter data set ($\tau_R = 0.4, \tau_T = 0.1$).

Then, we evaluated the effect of hierarchical hybrid signatures, which were judiciously selected to improve the performance of hybrid filtering. We compared filtering algorithms with hash-based hybrid signatures and hierarchical hybrid signatures given different constraints of index sizes (i.e., maximum numbers of signature elements as mentioned in Section 5.2). Figure 15 shows the experimental results. Hierarchical hybrid signatures achieved better performance compared with hash-based hybrid signatures in various index sizes. For example, in Figure 15(b), the elapsed time of algorithm with hierarchical hybrid signatures was 30 milliseconds given index size 280 MB, while that of hash-based hybrid signatures was 70 milliseconds. The main reason is that we judiciously selected hybrid signatures with hierarchical grids. These signatures improved the filtering power, and thus pruned a large number of dissimilar objects.

### 6.5 Comparison with Existing Methods

We compared our algorithm using hierarchical hybrid signatures (denoted by SEAL) with the keyword-first method (KEYWORD), the spatial-first method (SPATIAL), and state-of-the-art spatial keyword search method IR-tree [7] as discussed in Section 2.3. Figures 16 and 17 respectively show the experimental results on Twitter and USA datasets.

We can see that KEYWORD and SPATIAL could not effectively prune dissimilar objects. Specifically, KEYWORD sometimes performed worse than SPATIAL (See Figure 17(a)), as it did not have spatial pruning power. On the other hand, for large textual thresholds, SPATIAL might achieve lower efficiency (See Figures 16(d) and 17(d)), as it did not have textual pruning power. In addition, IR-tree also achieved low performance, and it was even worse than SPATIAL (See Figures 16). This is because the method, which is designed for spatial keyword search, visited too many unnecessary nodes and involved a huge number of dissimilar objects as mentioned in Section 2.3. In addition, IR-tree had to probe the inverted file associated in each R-tree node, resulting in a large overhead.

Our method SEAL always achieved the highest performance for any type of queries and any threshold on the two datasets. In Figures 16 and 17, our method was several tens of times faster than the baseline methods. For example, in Figure 16(c), given spatial threshold 0.1 and textual threshold 0.4, our method took 5 milliseconds, while IR-tree, KEYWORD and SPATIAL respectively took 253, 52 and 182 milliseconds. The better performance of our method is attributed to the signature-based methods integrating both spatial and textual pruning simultaneously, and the hierarchical hybrid signatures we selected for large filtering power.

### 6.6 Scalability

We evaluated the scalability of our hybrid filtering algorithm by varying the numbers of objects. Figure 18 shows the results. We can see that our method scaled very well, and with the increase of the numbers of objects, the elapsed time increased sub-linearly. This is because our algorithm could prune a huge amount of dissimilar objects even if the number of objects increases.

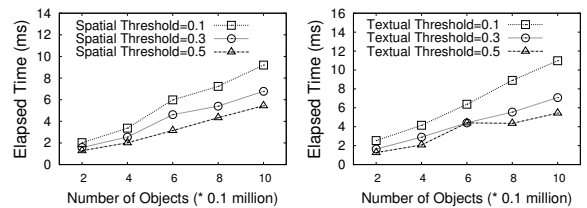

(a) Large-Region Queries.   (b) Large-Region Queries.
Figure 18: Scalability on the Twitter data set.

## 7. CONCLUSION AND FUTURE WORK

In this paper, we have studied a new research problem called spatio-textual similarity search. We introduced a filter-and-verification framework to compute the answers. We devised efficient signature-based filtering algorithms and developed effective pruning techniques. For spatial pruning, we proposed grid-based signatures by decomposing the underlying space, and developed threshold-aware pruning techniques. To utilize spatial and textual pruning simultaneously, we judiciously selected hierarchical hybrid signatures and devised hybrid filtering algorithms. We have implemented our method and examined it on real and synthetic datasets. Experimental results show that our method achieves very high search performance.

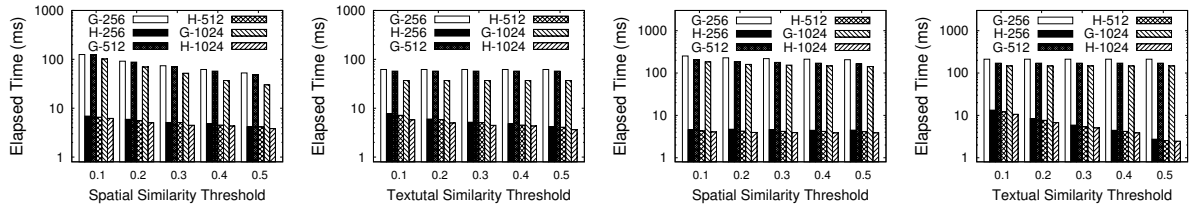

(a) Large-Region Query. (b) Large-Region Query. (c) Small-Region Query. (d) Small-Region Query.

Figure 14: Comparison of grid-based and hybrid filters on the Twitter data set.

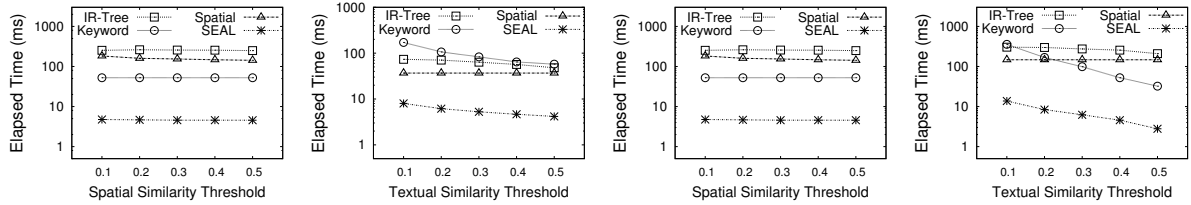

(a) Large-Region Query. (b) Large-Region Query. (c) Small-Region Query. (d) Small-Region Query.

Figure 16: Comparison with existing methods on the Twitter data set.

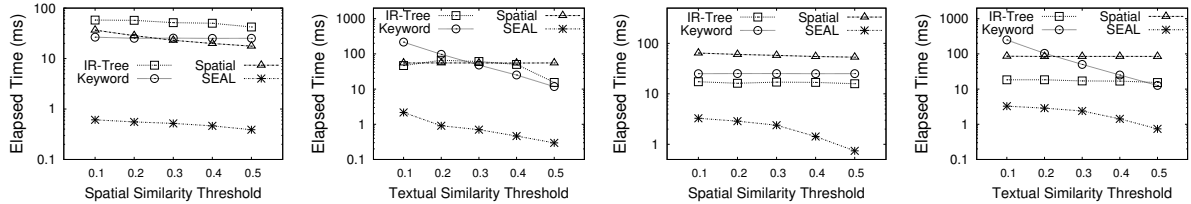

(a) Large-Region Query. (b) Large-Region Query. (c) Small-Region Query. (d) Small-Region Query.

Figure 17: Comparison with existing methods on the USA data set.

We believe this study on spatio-textual similarity search opens many new interesting and challenging problems that need further research investigation, such as how to extend the textual similarity measure to more sophisticated schemes, how to provide a theoretical analysis of the approximate solutions presented in this paper, etc.

**Acknowledgement.** This work was partly supported by the National Natural Science Foundation of China under Grant No. 61003004 and No. 60833003, the National Grand Fundamental Research 973 Program of China under Grant No. 2011CB302206, a project of Tsinghua University under Grant No. 20111081073, and the "NExT Research Center" funded by MDA, Singapore, under the Grant No. WBS:R-252-300-001-490.